\documentclass[pra,twocolumn]{revtex4}
\usepackage{graphicx}
\usepackage{amssymb}
\usepackage{amsmath}
\begin{document}
\title{Sub-Doppler Laser Cooling and Magnetic Trapping of Erbium}
\author{Andrew J. Berglund$^1$}
\author{Siu Au Lee$^2$}
\author{Jabez J. McClelland$^1$}\affiliation{$^1$Center for Nanoscale Science and Technology,
National Institute of Standards and Technology, Gaithersburg, MD
20899\\{$^2$Department of Physics, Colorado State University, Ft.
Collins, CO 80534}}
% PACS
% 32.80.Pj Optical cooling of atoms; trapping

\begin{abstract}We investigate cooling mechanisms in
magneto-optically and magnetically trapped erbium. We find efficient
sub-Doppler cooling in our trap, which can persist even in large
magnetic fields due to the near degeneracy of two Land\'{e} $g$
factors. Furthermore, a continuously loaded magnetic trap is
demonstrated where we observe temperatures below 25 $\mu$K. These
favorable cooling and trapping properties suggest a number of
scientific possibilities for rare-earth atomic physics, including
narrow linewidth laser cooling and spectroscopy, unique collision
studies, and degenerate bosonic and fermionic gases with long-range
magnetic dipole coupling.
\end{abstract}
\maketitle

\section{Introduction}

The maturation of laser cooling and trapping techniques has enabled
many recent advances with elements ranging over much of the periodic
table. Narrow intercombination lines in Sr provide ultra-precise
atomic frequency standards \cite{Takamoto:2003a,Boyd:2006a}; Yb has
also been proposed as a frequency standard
\cite{porsev:2004a:021403,hoyt:2005a:083003}, with quantum
degeneracy recently reached in both bosonic \cite{Takasu:2003a} and
fermionic isotopes \cite{fukuhara:2007a:030401}; a Ra
magneto-optical trap (MOT) promises new tests of fundamental
symmetries, while also exhibiting an interesting blackbody radiation
repumping mechanism \cite{guest:2007a:093001}; Cr has a large
ground-state magnetic moment (6$\mu_B$), which is expected to give
long-range order to a Bose-Einstein condensate
\cite{griesmaier:2005a:160401}. In an extension of laser cooling
techniques to a highly magnetic rare earth atom, one of us recently
reported an Er MOT where a novel recycling mechanism allows trapping
despite significant leakage into energy levels that are dark to the
cooling light \cite{mcclelland:2006a:143005}.

Along with the other rare-earth elements, erbium offers an exciting
combination of favorable properties for future cold atomic physics
experiments. All six stable isotopes have been trapped, with
significant populations in three bosonic ($^{166}$Er, $^{168}$Er,
$^{170}$Er) and one fermionic ($^{167}$Er, nuclear spin $I=7/2$)
species. Er exhibits a variety of accessible transitions that are
useful for spectroscopy or laser cooling, including a broad (36~MHz)
line at 401 nm, a narrow (8~kHz) line at 841 nm, and an ultra-narrow
(2~Hz) line in the telecommunication spectrum at 1299 nm
\cite{Martin:1978a,Ban:2005a,mcclelland:2006b:064502}. It has a very
large ground-state magnetic moment (7$\mu_B$), which produces a
strong magnetic dipole interaction at low temperatures. Its closed
$5s^2$ and $6s^2$ electron shells shield the $4f$ core electrons,
resulting in novel scattering properties while also suppressing
undesirable Zeeman relaxation rates \cite{hancox:2004a}.
Additionally, a deterministic ion implantation source based on an Er
MOT \cite{hanssen:063416} could provide a single-ion analog of a
solid-state laser while dramatically enhancing the prospects for
scalability in recent rare-earth quantum computing proposals
\cite{wesenberg:2007a:012304}. However, none of these avenues can be
pursued without first understanding the laser cooling and trapping
properties of this new system.

The primary result of this paper is an exhibition of cooling and
trapping properties, both magnetic and optical, which bolster the
growing list of erbium's desirable features for future cold atomic
physics. First, we show that efficient sub-Doppler cooling occurs in
our MOT where temperatures reach as low as 100~$\mu$K, nearly an
order of magnitude below the Doppler temperature ($T_D=860~\mu$K).
Second, we point out that for the 401~nm transition, the Land\'{e}
$g$ factors describing the Zeeman shift of the ground- and
excited-state magnetic sublevels are nearly equal. This near
degeneracy of Land\'{e} $g$ should soften the usual constraint that
$\sigma^+\sigma^-$ polarization gradient cooling can only occur in
negligible magnetic fields and may have interesting applications for
atomic beam collimation or atomic waveguiding along magnetic field
lines. Finally, we show that a magnetically trapped atom fraction in
our MOT is unexpectedly cold, with temperatures falling below 25
$\mu$K, lower than expectations based on simple thermodynamic
considerations. We believe these magneto-optical properties, coupled
with the unique atomic properties of the rare-earth elements, make
Er an exciting candidate for a wide range of future experiments.

\section{Sub-Doppler cooling in the MOT}
\begin{figure}[t]\includegraphics*[width=3.2in]{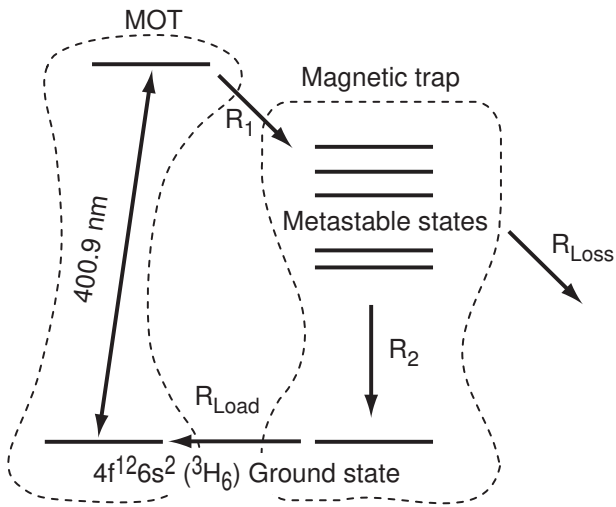}
\caption{\label{Fig:LevelDiagram}(Color online) Schematic level
diagram of the Er MOT and magnetic trap
\cite{mcclelland:2006a:143005}. The excited state of the 401~nm
transition decays into magnetically trapped metastable states at a
rate $R_1\approx 1700$~s$^{-1}$. Relaxation to the ground state at
rate $R_2$ and trap loss at rate $R_{\mathrm{Loss}}$ both occur with
rates of 4~s$^{-1}$ to 5~s$^{-1}$. $R_{\mathrm{Load}}$ varies
depending on MOT conditions.}\end{figure}

Our apparatus was described in \cite{mcclelland:2006a:143005}. An
effusive atomic beam is produced in the horizontal ($x$) direction
by heating a sample of Er to $1200^\circ$C in a crucible with a 1 mm
aperture. The beam is subsequently decelerated in a $\sigma^-$
Zeeman slower \cite{Barrett:1991a:PhysRevLett.67.3483} then loaded
into a MOT operating on the strong ($\Gamma/2\pi=36$~MHz) transition
at 401 nm. Cooling and trapping light is derived from a
frequency-doubled Ti:sapphire laser. The Zeeman slower beam, MOT
beams, and a probe beam are derived from the same laser, with
frequency offsets and relative intensities controlled by
acousto-optic modulators. All laser frequencies are calibrated and
locked to a fluorescence probe on the atomic beam, with an overall
uncertainty of $\pm 2.8$~MHz \footnote{Unless otherwise noted, all
uncertainties represent one standard deviation combined random and
systematic uncertainty}. The quadrupole magnetic field at the trap
position is the sum of fields from our MOT coils together with the
Zeeman slower and compensation coils, with the largest gradient
along the direction of gravity ($z$). Typical field gradients are
$(dB_x,dB_y,dB_z) = (0.20,0.11,-0.33)$~T/m with $\pm5\%$
uncertainty. All of the measurements described here were performed
on $^{166}$Er, with trapped atom populations between $10^4$ and
$2\times 10^5$.

\begin{figure}
\includegraphics*[width=3.2in]{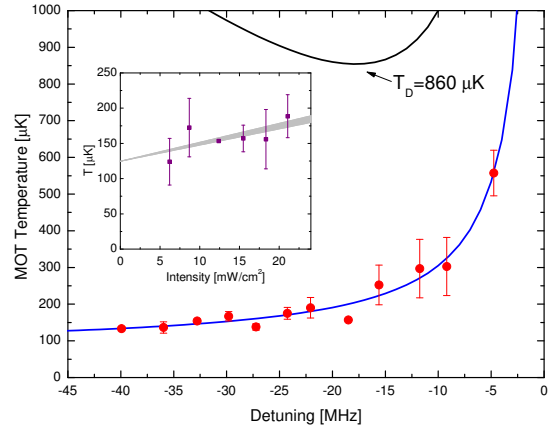}
\caption{\label{Fig:MOTTVsDetuning}(Color online) MOT temperature
vs. detuning for $I=14.3$~mW/cm$^2$. The upper curve is expected for
Doppler cooling a two-level atom, while the lower curve is a fit to
Eq.~\eqref{Eq:SubDopplerScaling}. The atom number ranged from a
minimum value of approximately $10^4$ at $\delta/2\pi = -5~$MHz to a
maximum of $1.3\times 10^5$ at $\delta/2\pi = -30~$MHz
$(-0.8\Gamma)$. (Inset) MOT temperature vs. trapping beam intensity
at a fixed detuning $\delta=-0.7\Gamma$. The shaded gray area is a
fit to Eq.~\eqref{Eq:SubDopplerScaling} together with the spread
resulting from uncertainty in $\delta$. Error bars are described in
the text.}\end{figure}

Er exhibits a complex level structure, with 110 electronic states
lying between the ground and excited state of the 401 nm
laser-cooling transition. As shown in Fig.~\ref{Fig:LevelDiagram},
the MOT remains functional because decay of excited atoms into
metastable intermediate states does not necessarily result in the
loss of those atoms. On the contrary, a large fraction of these
metastable, dark-state atoms remain magnetically trapped and
eventually relax back to the ground state. We begin by studying the
temperature of the MOT, which we measure by observing the ballistic
expansion of the atom cloud after all optical and magnetic fields
are extinguished. The cloud is imaged by pulsing the MOT beams for
200~$\mu$s and recording the resulting fluorescence image (in the
$xz$ plane) using a CCD camera. Fig.~\ref{Fig:MOTTVsDetuning} shows
the measured temperature $T$ as a function of the detuning of the
MOT beams, $\delta$. The error bars represent uncertainty estimates
from variation in the expansion velocity determined independently
along the $x$ and $z$ directions. The upper solid curve shows the
prediction of simple Doppler theory for a two-level atom
\cite{Metcalf:1999}, while the lower curve is a fit to a
characteristic sub-Doppler scaling law \cite{Drewsen:1994a,Xu:2003a}
\begin{equation}\label{Eq:SubDopplerScaling} T = T_0+C_{\sigma^+\sigma^-} \frac{ \hbar
\Gamma}{2 k_B}\left(\frac{\Gamma}{|\delta|}\right) \frac{I}{I_s}
\end{equation}where $I$ is the total intensity at the MOT
position and $I_s =72$~mW/cm$^2$ is the saturation intensity. The
resulting fit gives $C_{\sigma^+\sigma^-}~=~0.38\pm0.02$ and a
predicted minimum temperature $T_0=(77\pm10)~\mu$K in the
zero-intensity limit. We also varied the intensity at fixed detuning
$\delta=-0.7\Gamma$ (inset of Fig.~\ref{Fig:MOTTVsDetuning}), to
find $C_{\sigma^+\sigma^-}=0.15\pm0.08$ and $T_0=(125\pm21)\mu$K.
The difference in fit parameters between the data sets indicates
that the temperature may have a more complicated dependence on
intensity than the simple scaling law of
Eq.~\eqref{Eq:SubDopplerScaling}. This is not too surprising, since
in our trap, intensity-dependent rates of optical pumping into
magnetically trapped dark states may strongly affect the MOT
temperature (we show below that the magnetically trapped atom
fraction exhibits anomalous temperature behavior). Nevertheless, our
results show qualitatively and quantitatively that strong, efficient
sub-Doppler cooling occurs in the MOT. We emphasize that this is a
single-stage process (see also Ref.~\cite{Xu:2003a}), where
sub-Doppler cooling occurs in the MOT without an additional
``optical molasses" period of polarization-gradient cooling in a
nulled magnetic field.

\begin{figure}[t]\includegraphics*[width=3.2in]{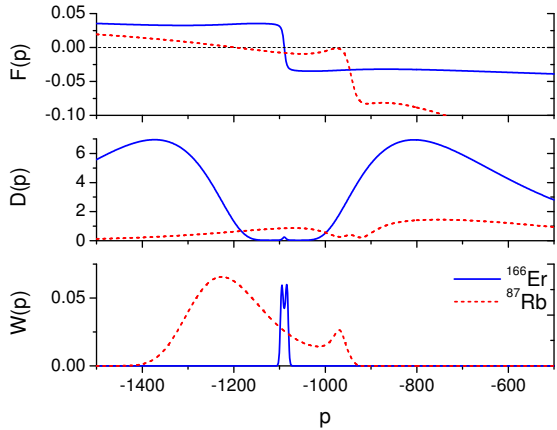} \caption{\label{Fig:SubDopplerCalcs}
(Color online) Semiclassical force $F(p)$ (in units of $\hbar
k\Gamma$), momentum diffusion coefficient $D(p)$ (in units of
$\hbar^2k^2\Gamma$), and equilibrium probability distribution $W(p)$
(arbitrary units) for $^{87}$Rb (dashed lines, red) and $^{166}$Er
(solid lines, blue) in a $\sigma^+\sigma^-$ optical molasses in a
longitudinal magnetic field $B=1$~mT. For both cases, the saturation
parameter $s=2$ and detuning $\delta=-2.5\Gamma$.}
\end{figure}

As discussed in Refs.~\cite{Walhout:1992a,Werner:1992a},
$\sigma^+\sigma^-$ polarization-gradient cooling typically breaks
down in a longitudinal magnetic field because the ``locking
velocity," where atomic-motion induced frequency shifts are balanced
by Zeeman shifts, is generally different for sub-Doppler and Doppler
cooling mechanisms. This difference arises because Doppler
mechanisms rely on absorption and spontaneous emission, so Zeeman
shifts of both the ground and excited states affect the cooling
efficiency; on the other hand, sub-Doppler mechanisms rely on
stimulated processes between ground state sub-levels, and thus
depend primarily on Zeeman shifts in the ground state. Because the
Land\'{e} $g$ factors of the ground and excited states of a cooling
transition are generally unequal, there is a critical magnetic field
value at which the two mechanisms ``unlock," with the larger capture
range of the Doppler cooling force dominating beyond this point
\cite{Walhout:1992a,Werner:1992a}. For most laser-cooled atoms, this
critical magnetic field value is around 0.1~mT. In Xe, for example,
the difference in $g$ factors between the ground and excited state
is $\Delta g_{eg}=-0.17$ and polarization-gradient cooling was
observed to fail at 0.2 mT~\cite{Walhout:1996a}. On the other hand,
the $g$ factor difference for the 401 nm transition in Er is $\Delta
g_{eg} = -0.004$ \cite{Martin:1978a}, suggesting that
$\sigma^+\sigma^-$ polarization-gradient cooling should remain
effective in longitudinal magnetic fields as large as 10~mT.

To investigate the possible effect of $g$ factor degeneracy on Er
sub-Doppler cooling in a magnetic field, we calculated the
semiclassical force $F(p)$, momentum diffusion coefficient $D(p)$,
and steady-state momentum distribution $W(p)$ for an atom with
momentum $\hbar k p$ in a one-dimensional $\sigma^+\sigma^-$ optical
molasses in a longitudinal magnetic field $B$. Our calculations
follow the method of Ref.~\cite{Walhout:1992a}, in which $p$ is
treated as a classical variable and a Fokker-Planck equation is
derived by expanding the optical Bloch equations to second order in
$\hbar k$. Results are shown in Fig.~\ref{Fig:SubDopplerCalcs} for
the 401~nm line of $^{166}$Er and for the D$_2$ line of $^{87}$Rb
($\Delta g_{eg}=-0.17$) in a magnetic field $B=1$~mT. Because of the
complicated interplay between optical coherences, Zeeman coherences,
and atomic momentum, the distributions $W(p)$ are clearly not
Gaussian. As a result, a temperature does not exist in the strict
sense, but we may still take the variances of $W(p)$ as a measure of
the momentum spread. These variances correspond to center-of-mass
``temperatures" of $25~\mu$K and $3.6$~mK for $^{166}$Er and
$^{87}$Rb, respectively. The same calculation gives respective
temperatures of $25~\mu$K and $17~\mu$K at zero magnetic field
$B=0$. For this simplified one-dimensional case, the $g$-factor
difference results in a temperature increase of more than 2 orders
of magnitude for $^{87}$Rb, with no change in temperature for
$^{166}$Er, at a magnetic field value of $B=1$~mT (10~G). Note that
this value of the magnetic field is attained in our setup at a
distance of a few millimeters from the MOT position, within the beam
waist of our trapping lasers. The near degeneracy of Er $g$-factors
may contribute to capture and loading processes in our MOT and
should also have interesting applications for laser cooling Er in
moderately large magnetic fields.

\begin{figure}[t]
  \centering\includegraphics*[width=3.2in]{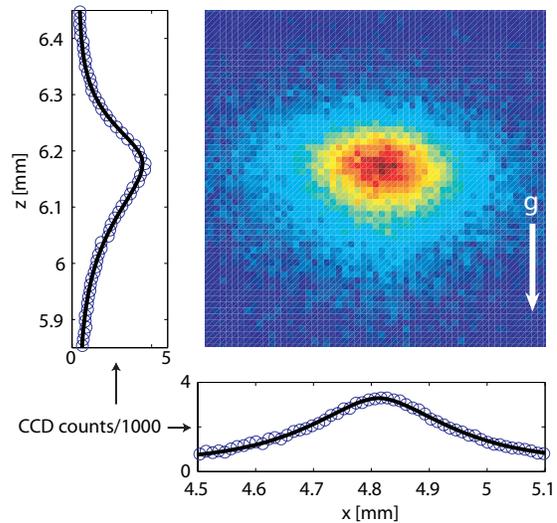}\caption{\label{Fig:BTrapImage}
  (Color online) CCD image of magnetically trapped atoms after 0.3~s in the magnetic trap. The marginal
  distributions and fits to $p_x(x)$ and $p_z(z)$ are shown at bottom and left, respectively.
  For this image, the fit parameters are $\bar{x}=150~\mu$m, $\bar{z}=105~\mu$m,
  and $\bar{g}=0.23$, corresponding to an effective magnetic moment
  $\bar{\mu}=3.9~\mu_B$ and temperature $T=(44\pm7)~\mu$K.}
\end{figure}
\section{Magnetic Trapping}

\begin{figure}[t]
  \includegraphics*[width=3.2in]{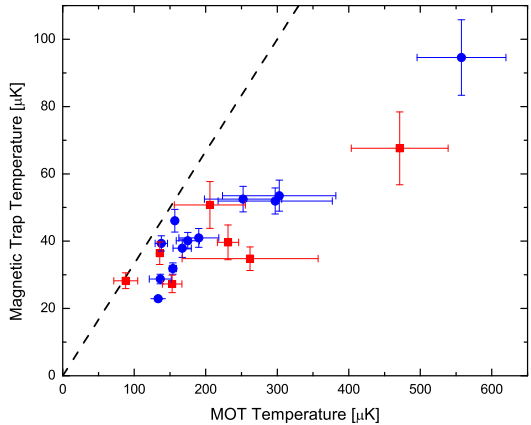}\caption{\label{Fig:MOTTvsBTrapT}
  (Color online) Magnetic trap temperature vs. MOT temperature. The MOT temperature was
  varied by changing the detuning $\delta$.
  Blue circles and red squares represent data from
  separate experimental runs at different Zeeman slower and MOT beam intensities.
  Simple thermodynamic arguments predict that all
  points should lie above the dashed line where $T_B\geq T_M/3$.}
\end{figure}
We now consider the fraction of ground-state atoms that remain
magnetically trapped after the MOT beams have been turned off.
Assuming these atoms can be described by an \emph{effective}
magnetic moment $\bar{\mu}\geq 0$, then the potential energy for a
trapped atom is given by
\begin{equation*}\label{Eq:MagneticEnergy}U(x,y,z) = \bar{\mu} \sqrt{dB_x^2 x^2 +
dB_y^2 y^2+dB_z^2z^2}+M g z
\end{equation*} where $M$ is the atomic mass, and $g$ is the
gravitational acceleration. In equilibrium at temperature $T$, the
marginal distributions of atoms along the $x$ and $z$ directions are
found to be
\begin{equation*}\begin{split} &p_x(x) =N_x\exp\left(-
2\frac{|x|}{\bar{x}}\sqrt{1-\bar{g}^2}\right)\left(1+2\frac{|x|}{\bar{x}}\sqrt{1-\bar{g}^2}\right)\\
&p_z(z) = N_z \exp\left(- 2\frac{|z|}{\bar{z}}-
2\bar{g}\frac{z}{\bar{z}}\right)\left(1+2\frac{|z|}{\bar{z}}\right)
\end{split}\end{equation*} where $N_x$ and $N_z$ are normalization constants.
The length scales and effects of gravity are characterized by
$\bar{g}~=~Mg/(\bar{\mu}dB_z)$ and $\bar{x}=2k_B T/(\bar{\mu}
dB_{x})$ with a similar expression for $\bar{z}$. Such a trap is
stable against gravity only if $\bar{g}<1$. For $\bar{g}=0$,
$\bar{x}$ and $\bar{z}$ are the standard deviations of the cloud,
while for $\bar{g}>0$, the cloud stretches along gravity. We capture
images of magnetically trapped atomic clouds by loading the MOT for
2~s then extinguishing the loading and trapping lasers, leaving all
magnetic fields unchanged. After a variable hold time $\Delta t$, we
pulse the MOT beams and capture the resulting fluorescence image.
Such an image is shown in Fig.~\ref{Fig:BTrapImage}. By analyzing
these images, we can determine $T$, $\bar{\mu}$, and atom number as
the magnetic trap evolves in time \footnote{In our image analysis,
we do not account for Zeeman shifts over the spatial extent of the
cloud, which are expected to alter the excited state fraction by at
most 7\%.}.

\begin{figure}[t]
\includegraphics*[width=3.2in]{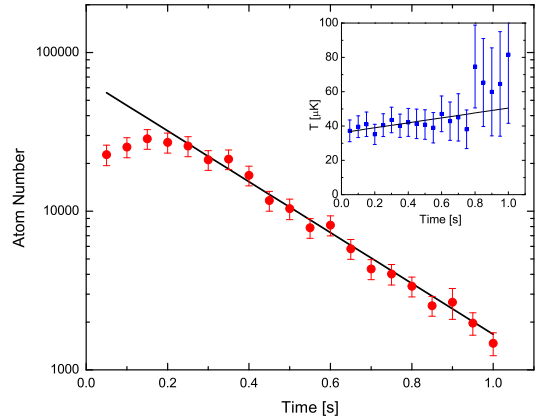}
\caption{\label{Fig:BTrapDecay}(Color online) Evolution of the
magnetically trapped atom number. After a time $\Delta t\approx
0.2$~s, all of the metastable atoms have returned to the ground
state and the trap decays exponentially with a time constant of
$(270\pm13)$~ms (solid line). (Inset) The temperature of the trapped
atoms increases over time with an error-weighted heating rate of
$(14\pm4)~\mu$K/s (solid line).}
\end{figure}
If the magnetic trap loading efficiency is independent of an atom's
energy, then thermodynamic arguments
\cite{Stuhler:2001a:PhysRevA.64.031405} predict that the
magnetically trapped atom temperature $T_B$ should be related to the
MOT temperature $T_M$ by $T_B\geq T_M/3$ (equality holds when the
spatial extent of the MOT is much smaller than that of the magnetic
trap). However, our magnetic trap temperatures systematically
violate this bound as shown in Fig.~\ref{Fig:MOTTvsBTrapT} where we
measure temperatures below 25~$\mu$K. Similar anomalously low
temperatures have been observed in MOT-loaded, magnetically trapped
Ca \cite{Hansen:2003a} and Sr \cite{Nagel:2003a} while unexpectedly
high temperatures were reported in Cr
\cite{Stuhler:2001a:PhysRevA.64.031405}. In Ref. \cite{Nagel:2003a},
a selection effect during trap loading was suggested to explain the
low temperatures. We suspect a similar phenomenon is at work here,
in which spatially-dependent optical pumping in the MOT tends to
correlate an atom's spin state with its energy; the ejection of
atoms in untrapped magnetic states ($m_J\leq 0$) may then reduce the
average energy of the sample when the MOT is turned off. This
hypothesis cannot be quantitatively evaluated without studying the
loading efficiency of the magnetic trap, which is difficult to
assess due to the presence of trapped, dark-state atoms that cannot
be probed directly.

An example of the time evolution of the magnetic trap population and
temperature is shown in Fig.~\ref{Fig:BTrapDecay}. Before $\Delta
t\approx 0.2~$s, the trap population appears to increase. This is
due to the relaxation of metastable atoms back to the ground
electronic state where they are resonant with the probe light (MOT
beams). After this time, we see single-exponential decay, most
likely arising from collisions with the background gas, with a
pressure around $2.3\times10^{-6}$~Pa \footnote{We find good fits to
the magnetic trap relaxation data of Fig.~\ref{Fig:BTrapDecay} for a
simple model in which a fraction $f$ of initially dark (metastable)
atoms relaxes to the ground state exponentially while atoms are
simultaneously lost from the trap at a second exponential rate.
However, this larger number of fit parameters cannot give an
independent indication of both the initial dark state fraction and
the metastable relaxation rate so we have only included a fit to the
long-time relaxation behavior here.}. This claim is further
supported by a strong correlation between background gas pressure
and trap lifetime and the similar lifetimes for $^{166}$Er and
$^{168}$Er (data not shown). We do not suspect that the trap decay
rate is due to density-dependent collisional effects within the
trap, because the number densities in the present measurements are
not particularly large [we observe peak densities of
$n=(2.1\pm0.3)\times 10^{9}$~cm$^{-3}$] and because we see no
evidence of non-exponential decay. On the other hand, we do not know
the origin of the observed heating rate, which may arise from
dipolar relaxation processes or collisional heating mechanisms.
Improvements to the vacuum system should eliminate our dominant loss
channel and allow us to study other relaxation processes. Finally,
our data indicate a phase-space density $\rho\approx
2.5\times10^{-8}$. It should be straightforward to increase this by
at least an order of magnitude by increasing the atomic beam flux.

\section{Conclusions}
In summary, we have presented a number of favorable cooling and
trapping properties for atomic erbium. We showed that efficient
sub-Doppler cooling occurs in our MOT without the need for an
additional optical molasses cooling stage. Furthermore, sub-Doppler
cooling should remain effective in large magnetic fields due to the
near degeneracy of Land\'{e} $g$ factors. This phenomenon may have
interesting applications for laser cooling in large magnetic
gradient traps, and might be useful for constructing atomic
waveguides along magnetic field lines. We also studied the
thermodynamics of a continuously-loaded magnetic trap, where
temperatures reach unexpectedly low values. The efficient cooling
and trapping methods discussed here should be straightforwardly
useful as a starting point for phase-space compression towards
quantum degeneracy. However, magnetic and collisional relaxation
processes must be studied before such a procedure can be considered.

\acknowledgements{The authors gratefully acknowledge J.~L. Hanssen,
N. Lundblad, and J.~V. Porto for stimulating discussions and D.
Rutter and A. Band for technical assistance. S.~L. acknowledges
support from NIST and NSF. A.~B. received financial support from the
National Research Council.}

%\bibliography{ajb_bib}

\end{document}